\documentclass[twocolumn,showpacs,preprintnumbers,pra,amsmath,amssymb]{revtex4}

\usepackage{graphicx}
\usepackage{bm}
\begin{document}
\title{Improved Measurement-Device-Independent Quantum Key Distribution with Uncharacterized Qubits }

\author{Won-Young Hwang$^{1}$\footnote{Email: wyhwang@jnu.ac.kr}, Hong-Yi Su$^{1}$, and Joonwoo Bae$^{2}$}

\affiliation{$^{1}$ Department of Physics Education, Chonnam National University, Gwangju 61186, Republic of Korea\\
$^{2}$ Department of Applied Mathematics, Hanyang University (ERICA), Ansan, Gyeonggi-do 15588, Republic of Korea.
}
\begin{abstract}
We propose an improved bound for the difference between phase and bit error rate in measurement-device-independent quantum key distribution with uncharacterized qubits. We show by simulations that the bound considerably increases the final key rates.
\end{abstract}
\maketitle
\section{Introduction}
Quantum key distribution (QKD) \cite{Sca09} enables two remote users Alice and Bob to generate key, which is not a classically possible task. QKD is not only a practically important field but also a theoretically appealing one. After security of QKD for ideal devices was shown \cite{May01,Sho00}, problems due to imperfect devices surfaced. Although a problem due to imperfect source was resolved \cite{Hwa03}, problem due to imperfect detectors still had remained. Device-independent (DI) QKD's elegantly overcome the imperfect device problems \cite{Aci07}.
However DI QKD is not yet feasible. In this background measurement-device-independent (MDI) QKD was proposed \cite{Lo12}. MDI QKD is secure provided that source is ideal, that is, source is exactly in the prescribed quantum states.
The latest protocols \cite{Yin13,Yin14} adapts a relaxed condition that source should be within $2$-dimensional subspace. In the protocols, bit error rate can be directly measured. But phase error rate is indirectly obtained by bounding difference between phase and bit error rates. Here phase error rate is more or less overestimated and thus key rate is reduced compared with previous MDI QKD's with characterized sources.

In this paper, we propose an improved bound for the difference between phase error rate and bit error rate. The improved one reduces the overestimation and thus gives larger key rates.

In section II, we briefly introduce the MDI QKD with uncharacterized qubits. In section III, we provide improved bound. In section VI, we discuss and conclude.
\section{MDI QKD with uncharacterized qubits}
For the protocol \cite{Yin14}, each user prepares two encoding states. Let the states prepared by Alice and Bob denoted by $|\varphi_m\rangle$ and $|\varphi_n^{\prime}\rangle$, respectively, where $m,n=0,1$. Here nothing is supposed about the encoding states so they are completely un-characterized. Each user also prepares a checking state {\it which is assumed to be a superposition of the encoding states}. Alice's and Bob's checking states are, respectively,
\begin{eqnarray}
|\varphi_2\rangle= c_0|\varphi_0\rangle + c_1 e^{i\theta}|\varphi_1\rangle,
 \nonumber
 \\
|\varphi_2^{\prime}\rangle= c_0^{\prime}|\varphi_0^{\prime}\rangle + c_1^{\prime} e^{i\theta^{\prime}}|\varphi_1^{\prime}\rangle.
\label{1}
\end{eqnarray}
Here $c_m$ and $c_n^{\prime}$ are non-negative numbers and $\theta$ and $\theta^{\prime}$ are real. The protocol is as follows.

(1) Alice generates a random number $i$ where $i=0,1,2$. She sends a state $|\varphi_i\rangle$ to Charlie. Here Charlie can be anyone. So Charlie can be either Eve (eavesdropper) or users themselves. (2) Bob independently generates a random number $j$ where $j=0,1,2$. He also sends a state $|\varphi_{j}^{\prime}\rangle$ to Charlie. (3) Charlie performs a measurement on the states $|\varphi_i\rangle$ and $|\varphi_{j}^{\prime}\rangle$. The measurement can be any one which finally gives two outcomes $0$ and $1$. Charlie announces the outcome.
(4) When the outcome is $0$, users discard the data. Otherwise, they keep the data. By sacrificing some of the data for public discussion, users estimate, $p_{ij}^1$, conditional probability to get outcome $1$ for each $i,j$.
(5) Measurement data with both $i$ and $j$ less than $2$ become raw key. Other data with either $i$ or $j$ is $2$ are used for checking purposes. Users do post-processing to get final key.

Now let us consider Charlie's measurement on the states $|\varphi_i\rangle$ and $|\varphi_{j}^{\prime}\rangle$. In the most general collective attack, Eve attaches an ancilla $|e\rangle$ to the states and then applies a unitary operation to them \cite{Yin14}
\begin{eqnarray}
 && U_{Eve}|\varphi_i\rangle |\varphi_j^{\prime}\rangle |e\rangle |0\rangle_M = \nonumber\\
 && \sqrt{p^0_{ij}}\hspace{1mm} |\Gamma_{ij}^0\rangle |0\rangle_M+  \sqrt{p^1_{ij}}\hspace{1mm} |\Gamma_{ij}^1\rangle |1\rangle_M.
\label{2}
\end{eqnarray}
Eve gets the outcome by measuring the quantum state indexed by $M$ in basis of $|0\rangle$ and $|1\rangle$. It is enough to consider only the data with the measurement outcome $1$ because those with $0$ is discarded. For convenience, let us omit $1$ from now on, $|\Gamma_{ij}^1\rangle \equiv |\Gamma_{ij}\rangle$ and $p_{ij}^1\equiv p_{ij}$. Now we can see that Eqs. (\ref{1}) and (\ref{2}) give constraints
\begin{eqnarray}
\sqrt{p_{2n}}\hspace{1mm} |\Gamma_{2n}\rangle &=& \sqrt{p_{0n}}\hspace{1mm} c_0|\Gamma_{0n} \rangle+ \sqrt{p_{1n}}\hspace{1mm} c_1 e^{i\theta} |\Gamma_{1n} \rangle,
\label{3}
\\
\sqrt{p_{m2}}\hspace{1mm} |\Gamma_{m2}\rangle &=& \sqrt{p_{m1}}\hspace{1mm} c_0^{\prime}|\Gamma_{m0} \rangle+ \sqrt{p_{m1}}\hspace{1mm} c_1^{\prime} e^{i\theta^{\prime}} |\Gamma_{m1} \rangle,
\label{4}
\\
\sqrt{p_{22}} \hspace{1mm} |\Gamma_{22}\rangle &=& \sum_{m,n} \sqrt{p_{mn}} \hspace{1mm} c_m c_n^{\prime} e^{i\theta_m} e^{i\theta_n^{\prime}}|\Gamma_{mn}\rangle,
\label{5}
\end{eqnarray}
where $\theta_0=\theta_0^{\prime}=0$, $\theta_1=\theta$, and
$\theta_1^{\prime}=\theta^{\prime}$. Constraints (\ref{3})-(\ref{5}) play key roles in security analysis.

For theoretical purposes, we consider an equivalent entanglement distillation protocol \cite{Sho00,Nie00}. In the hypothetical protocol, Alice and Bob prepare entangled states $(1/\sqrt{2})(|0\rangle_{A_1} |\varphi_0\rangle_{A_2} +|1\rangle_{A_1} |\varphi_1\rangle_{A_2})$ and  $(1/\sqrt{2})(|0\rangle_{B_1} |\varphi_0^{\prime}\rangle_{B_2} +|1\rangle_{B_1} |\varphi_1^{\prime}\rangle_{B_2})$, respectively. Alice and Bob send quantum states indexed by $A_2$ and $B_2$ to Charlie, respectively. According to Charlie's announcement about when measurement outcome is $1$, the users post-select their qubits indexed by $A_1$ and $B_1$, respectively. The post-selected state is given by \cite{Yin14}
\begin{eqnarray}
&&\rho= \frac{1}{p_{00}+p_{11}+p_{01}+p_{10}} \cdot \nonumber\\
&&\sum_q \bold{P}[\sqrt{p_{00}} \gamma_{00}^q |0\rangle_{A_1}|0\rangle_{B_1}+ \sqrt{p_{11}} \gamma_{11}^q |1\rangle_{A_1}|1\rangle_{B_1} \nonumber\\
&&+\sqrt{p_{01}} \gamma_{01}^q |0\rangle_{A_1}|1\rangle_{B_1}+ \sqrt{p_{10}} \gamma_{10}^q |1\rangle_{A_1}|0\rangle_{B_1}].
\label{6}
\end{eqnarray}
Here $\bold{P}[x]\equiv |x\rangle \langle x|$, $|\Gamma_{mn}\rangle \equiv \sum_{q} \gamma_{mn}^q |q\rangle$, where $|q\rangle$ are a set of orthonormal states. From normalization, $\sum_q |\gamma_{mn}^q|^2=1$. We consider four Bell states $|\varphi^{\pm \alpha} \rangle= (1/\sqrt{2})(|00\rangle \pm e^{i(\alpha_A+ \alpha_B)}|11\rangle)$ and $|\psi^{\pm \alpha} \rangle= (1/\sqrt{2})(|01\rangle \pm e^{i(\alpha_A- \alpha_B)}|10\rangle)$ which are obtained by rotating each qubit by Pauli operator $\sigma_z$ with amount $\alpha_A$ and $\alpha_B$, respectively. Final key rate is given by $R= 1-H(e_b)-H(e_p)$, where $H(x)= -x\log_2 x- (1-x)\log_2 (1-x)$ is binary Shannon entropy and $e_b$ and $e_p$ are bit and phase error rates, respectively.
 Bit error rate is given by
\begin{eqnarray}
 e_b&=&\langle \psi^{+\alpha}| \rho |\psi^{+\alpha} \rangle +
     \langle \psi^{-\alpha}| \rho |\psi^{-\alpha} \rangle
     \nonumber\\
    &=&\frac{p_{01}+p_{10}}{p_{00}+p_{11}+p_{01}+p_{10}}
\label{7}
\end{eqnarray}
for arbitrary $\alpha_A$ and $\alpha_B$.
Phase error rate is given by
\begin{eqnarray}
 e_p&=&\langle \varphi^{-\alpha}|\rho|\varphi^{-\alpha}\rangle +
       \langle \psi^{-\alpha}| \rho |\psi^{-\alpha} \rangle
       \nonumber\\
    &\leq& e_b+ \frac{\sum_q|\sqrt{p_{00}}\gamma_{00}^q- e^{i(\alpha_A+ \alpha_B)} \sqrt{p_{11}} \gamma_{11}^q|^2}{2(p_{00}+p_{11}+p_{01}+p_{10})}.
\label{8}
\end{eqnarray}
On the other hand, using normalization condition of the states in Eq. (\ref{1}) and constraints (\ref{3})-(\ref{5}), a term
$\sum_q|\sqrt{p_{00}}\gamma_{00}^q+ e^{i(\theta+ \theta^{\prime})} \sqrt{p_{11}} \gamma_{11}^q|^2$
can be upperbounded.
The normalization condition gives
\begin{eqnarray}
 (c_0 -c_1)^2 \leq 1 \leq (c_0 +c_1)^2, \nonumber\\
 (c_0^{\prime} -c_1^{\prime})^2
 \leq 1 \leq (c_0^{\prime} +c_1^{\prime})^2.
\label{A0}
\end{eqnarray}
Constraints (\ref{3}) and (\ref{4}) imply
\begin{eqnarray}
 (\sqrt{p_{00}}c_0 -\sqrt{p_{10}}c_1)^2
 \leq p_{20} \leq
 (\sqrt{p_{00}}c_0 +\sqrt{p_{10}}c_1)^2, \nonumber\\
 (\sqrt{p_{01}}c_0 -\sqrt{p_{11}}c_1)^2
 \leq p_{21} \leq
 (\sqrt{p_{01}}c_0 +\sqrt{p_{11}}c_1)^2, \nonumber\\
 (\sqrt{p_{00}}c_0^{\prime} -\sqrt{p_{01}}c_1^{\prime})^2
 \leq p_{02} \leq
 (\sqrt{p_{00}}c_0^{\prime} +\sqrt{p_{01}}c_1^{\prime})^2, \nonumber\\
 (\sqrt{p_{10}}c_0^{\prime} -\sqrt{p_{11}}c_1^{\prime})^2
 \leq p_{12} \leq
 (\sqrt{p_{10}}c_0^{\prime} +\sqrt{p_{11}}c_1^{\prime})^2,
\label{A1}
\end{eqnarray}
where $\sqrt{p_{mn}}$'s are fixed and $c_m$ are variables.

Constraint (\ref{5}) gives upperbound for $\sum_q|\sqrt{p_{00}}\gamma_{00}^q+ e^{i(\theta+ \theta^{\prime})} \sqrt{p_{11}} \gamma_{11}^q|^2$ \cite{Yin14},
\begin{eqnarray}
  \sum_q|\sqrt{p_{00}}\gamma_{00}^q+ e^{i(\theta+ \theta^{\prime})} \sqrt{p_{11}} \gamma_{11}^q|^2 \leq \max_{c,c^{\prime}} f(c,c^{\prime})
\label{A5}
\end{eqnarray}
where
\begin{eqnarray}
&&f(c,c^{\prime})= \min \nonumber\\
&&\{\frac{(\sqrt{p_{22}}+\sqrt{p_{01}} c_0 c_1^{\prime}
+\sqrt{p_{10}} c_1 c_0^{\prime}+\sqrt{p_{11}}| c_0 c_0^{\prime}- c_1 c_1^{\prime}|)^2}{(c_0 c_0^{\prime})^2}, \nonumber\\
&& \frac{(\sqrt{p_{22}}+\sqrt{p_{01}} c_0 c_1^{\prime}
+\sqrt{p_{10}} c_1 c_0^{\prime}+\sqrt{p_{00}}| c_0 c_0^{\prime}- c_1 c_1^{\prime}|)^2}{(c_1 c_1^{\prime})^2}\}.\nonumber\\
&&
\label{A6}
\end{eqnarray}
Optimization in Inequality (\ref{A5}) with constraints (\ref{A0}),(\ref{A1}) can be done numerically. According to our computer calculations, when bit error rate is small enough such that final key rate is nonnegligible, $c_0 c_0^{\prime}$ and $c_1 c_1^{\prime}$ don't become $0$. Thus here we consider only the case when the $c's$ are not zero.

The fact that the term $\sum_q|\sqrt{p_{00}}\gamma_{00}^q+ e^{i(\theta+ \theta^{\prime})} \sqrt{p_{11}} \gamma_{11}^q|^2$
has an upperbound means that $\Delta \equiv \sum_q|\sqrt{p_{00}}\gamma_{00}^q- e^{i(\alpha_A+ \alpha_B)} \sqrt{p_{11}} \gamma_{11}^q|^2$ has the same upperbound when $\alpha_A+ \alpha_B= \theta+ \theta^{\prime}+\pi$. Hence we can obtain upperbound for phase error rate, which then, with bit error rate, gives the final key rate.
\section{Improved bound for phase error rate}
Let us consider the quantity to be bounded. We get
\begin{eqnarray}
&&\Delta = \sum_q|\sqrt{p_{00}}\gamma_{00}^q- e^{i(\alpha_A+ \alpha_B)} \sqrt{p_{11}} \gamma_{11}^q|^2 \nonumber\\
&=& p_{00}+p_{11}-2\sqrt{p_{00}}\sqrt{p_{11}}\hspace{1mm}
\Re\hspace{0.5mm}[e^{i(\alpha_A+ \alpha_B)}\langle\Gamma_{00}|\Gamma_{11}\rangle ], \nonumber\\
\label{9}
\end{eqnarray}
where $\Re\hspace{0.5mm}[z]$ is real part of a complex number $z$. Constraint (\ref{5}) can be rewritten as
\begin{eqnarray}
&&\sqrt{p_{00}} \hspace{1mm} c_0 c_0^{\prime}|\Gamma_{00}\rangle
+\sqrt{p_{11}} \hspace{1mm} c_1 c_1^{\prime} e^{i(\theta+ \theta^{\prime})}|\Gamma_{11}\rangle = \nonumber\\
&&\sqrt{p_{22}} \hspace{1mm}|\Gamma_{22}\rangle-
\sqrt{p_{01}} \hspace{1mm} c_0 c_1^{\prime} e^{i\theta^{\prime}}|\Gamma_{01}\rangle-
\sqrt{p_{10}} \hspace{1mm} c_1 c_0^{\prime} e^{i\theta}|\Gamma_{10}\rangle,
\nonumber\\
\label{10}
\end{eqnarray}
from which we get
\begin{eqnarray}
&&\Re\hspace{0.5mm}[e^{i(\theta+\theta^{\prime})}
\langle\Gamma_{00}|\Gamma_{11}\rangle] \leq
\nonumber\\
&&
\frac{|\sqrt{p_{22}}+\sqrt{p_{01}} c_0 c_1^{\prime}
+\sqrt{p_{10}} c_1 c_0^{\prime}|^2-p_{00} c_0^2 {c_0^{\prime}}^2- p_{11} c_1^2 {c_1^{\prime}}^2}{2\sqrt{p_{00}}\sqrt{p_{11}} c_0 c_0^{\prime} c_1 c_1^{\prime}}
\nonumber\\
\label{11}
\end{eqnarray}
by $||\hspace{0.5mm}|A\rangle +|B\rangle|| \leq ||\hspace{0.5mm}|A\rangle||+||\hspace{0.5mm}|B\rangle ||$ where $|\hspace{0.5mm}A\rangle$, $|\hspace{0.5mm}B\rangle$ are arbitrary states and $||\hspace{0.5mm}|A\rangle||$ is norm of $|A\rangle$.

First we upperbound $\Re\hspace{0.5mm}[e^{i(\theta+\theta^{\prime})} \langle\Gamma_{00}|\Gamma_{11}\rangle]$ with constraints (\ref{3}) and (\ref{4}).
\begin{eqnarray}
&&\Re\hspace{0.5mm}[e^{i(\theta+\theta^{\prime})}
\langle\Gamma_{00}|\Gamma_{11}\rangle] \leq \max_{c,c^{\prime}}
\nonumber\\
&&
 \{\frac{|\sqrt{p_{22}}+\sqrt{p_{01}} c_0 c_1^{\prime}
+\sqrt{p_{10}} c_1 c_0^{\prime}|^2-p_{00} c_0^2 {c_0^{\prime}}^2- p_{11} c_1^2 {c_1^{\prime}}^2}{2\sqrt{p_{00}}\sqrt{p_{11}} c_0 c_0^{\prime} c_1 c_1^{\prime}}\}.
\nonumber\\
\label{12}
\end{eqnarray}
Maximization in Inequality (\ref{12}) can be done similarly with the constraints (\ref{A0}),(\ref{A1}).

Consider a case when the upperbound is negative, that is $\Re\hspace{0.5mm}[e^{i(\theta+\theta^{\prime})} \langle\Gamma_{00}|\Gamma_{11}\rangle]\leq -|\Omega|$, then we can see that $|\langle\Gamma_{00}|\Gamma_{11}\rangle|$ is lowerbounded by the absolute value of the upperbound,
$|\langle\Gamma_{00}|\Gamma_{11}\rangle| \geq |\Omega|$.
Now it is easy to see that we can always choose $\alpha_A+ \alpha_B$ such that
\begin{equation}
\Delta \leq p_{00}+ p_{11}- 2 \sqrt{p_{00}} \sqrt{p_{11}} \hspace{0.5mm} |\Omega|.
\label{19}
\end{equation}
Now we can get upperbound for phase error rate by Inequalities (\ref{8}) and (\ref{19}), from which we can get final key rate.

Let us compare the two upperbounds. First consider a case when
$p_{00}=p_{11}=(1/2)(1-e_b), p_{10}=p_{01}=(1/2)e_b, p_{20}=p_{21}=p_{02}=p_{12}= 1/4$, and $p_{22}= (1/2)e_b$.
The set of probabilities can be obtained by choosing that  $|\varphi_0\rangle =|0\rangle, |\varphi_1\rangle =|1\rangle, |\varphi_0^{\prime}\rangle=|0\rangle, |\varphi_1^{\prime}\rangle=|1\rangle$,
$|\varphi_2\rangle= (1/\sqrt{2})(|0\rangle+|1\rangle), |\varphi_2^{\prime}\rangle= (1/\sqrt{2})(|0\rangle-|1\rangle)$, and measurement element for outcome $1$ is $ |\varphi^{+}\rangle \langle \varphi^{+}|$. (Here we are not supposing that the measurements are done on the states actually. We consider the states and measurements in order to show that the set of probabilities are physically realizable. In the MDI QKD with uncharacterized qubits, security is shown regardless of how the set was generated.) We suppose channel between Alice and Charlie is depolarized \cite{Nie00}. But we  suppose channel between Bob and Charlie is perfect because the measurement unit is controlled by Bob. Here the higher the depolarization probability is, the higher the bit error rate is.
The final key rate and the quantity $\Delta/[2(p_{00}+p_{11}+p_{01}+p_{10}$)] are shown in Fig-1 and Fig-2.
\begin{figure}
\includegraphics[width=8cm]{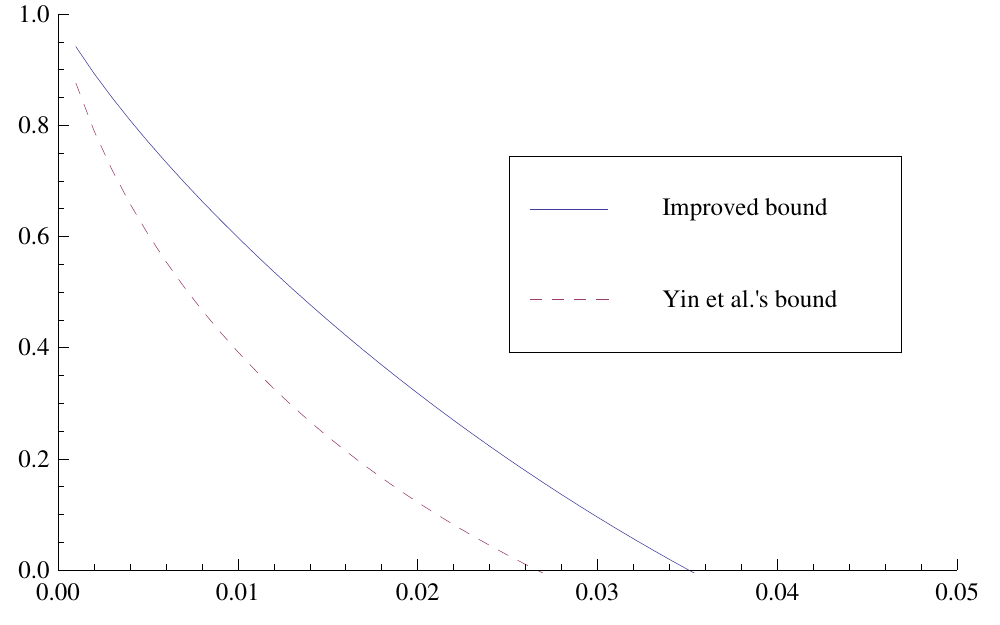}
\caption{Final key rate ($y$ axis) versus bit error rate ($x$ axis) for depolarization channel.}
\label{Fig-1}
\end{figure}
\begin{figure}
\includegraphics[width=8cm]{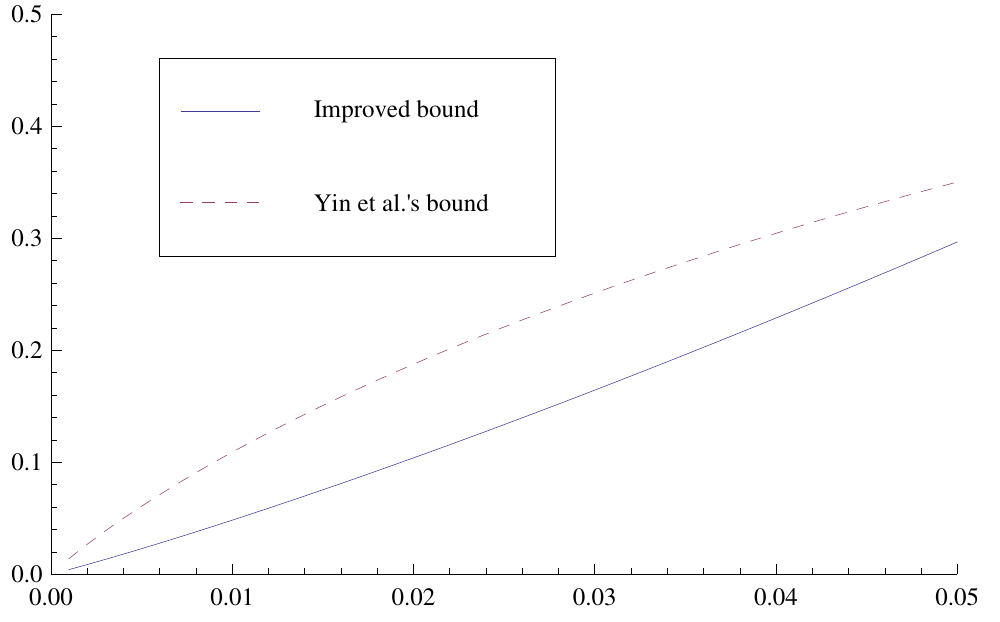}
\caption{$\Delta/[2(p_{00}+p_{11}+p_{01}+p_{10})]$($y$ axis) versus bit error rate ($x$ axis) for depolarization channel. $\Delta/[2(p_{00}+p_{11}+p_{01}+p_{10})]$ is bound for gap between phase and bit error rates.}
\label{Fig-2}
\end{figure}
Bit error rate that MDI QKD with uncharacterized qubit can tolerate is not that high as we can see in Fig-1. By the new bound, the tolerable bit error rate and the final key rate have been considerably improved.

Next let us consider the case \cite{Yin14} when four single-photon detectors with dark counting rate of $d= 10^{-5}$ are used in the measurement unit. Here $\eta$ denotes total transmission rate of the channel from Alice (Bob) to the measurement unit in the middle of channel and no Eve's attack is assumed. The same states used in the depolarization case are chosen. Now we have  $p_{00}=p_{11}= \eta^2(1-d)^2/2+ 2\eta(1-\eta)d(1-d)^2+ 2(1-\eta)^2 d^2(1-d)^2, p_{10}=p_{01}=p_{22}= 2\eta(1-\eta)d(1-d)^2+ 2(1-\eta)^2 d^2(1-d)^2$, and $p_{20}=p_{21}=p_{02}=p_{12}= (p_{00}+p_{01})/2$. The key rates by the two bounds and the original protocol \cite{Lo12} are described in Fig-3.
\begin{figure}
\includegraphics[width=8cm]{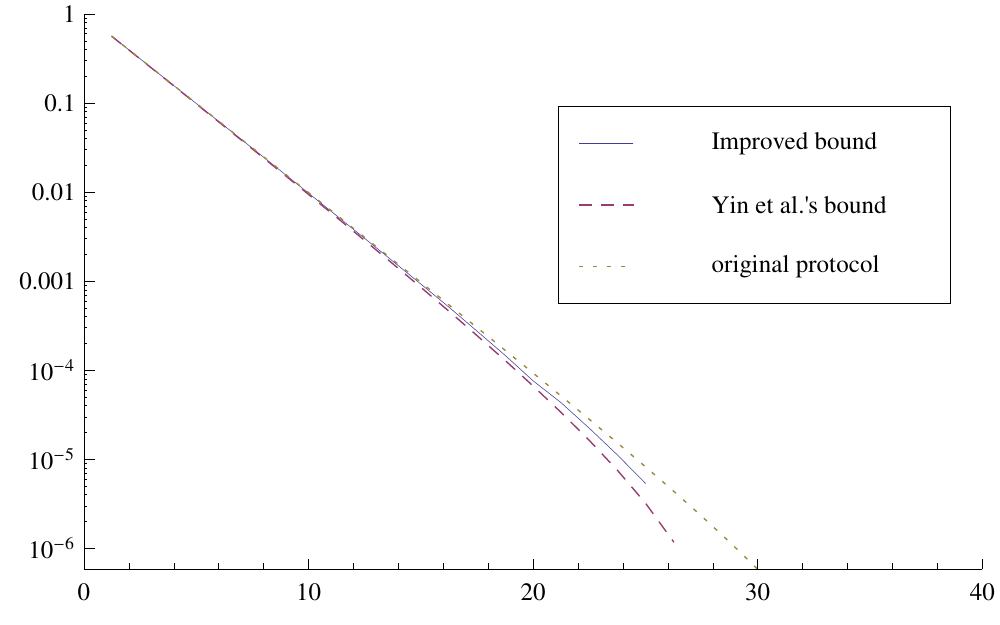}
\caption{Final key rate multiplied by $p_{00}+p_{11}+p_{01}+p_{10}$ ($y$ axis) versus total transmission loss by dB ($x$ axis) for four single-photon detectors with dark counting. In the original protocol, ideal source is supposed and thus phase error rate is equal to bit error rate.}
\label{Fig-3}
\end{figure}
Here the reason why the key rates do not differ much is that the bit error rate is not large.

\section{Discussion and Conclusion}
Inequality in (\ref{19}) implies that if $p_{00}=p_{11}$ and $|\langle\Gamma_{00}|\Gamma_{11}\rangle|=1$ then the $\Delta=0$ or the phase error rate equals to bit error rate for a certain choice of Bell states. This can be confirmed by observing that the state $|\Psi\rangle \langle \Psi|= \rho$ in Eq. (\ref{6}) becomes $|\Psi\rangle= \sqrt{p_{00}} (|00\rangle+ e^{iA}|11\rangle) |\Gamma_{00}\rangle + \sqrt{p_{01}} |01\rangle |\Gamma_{01}\rangle +\sqrt{p_{10}} |10\rangle |\Gamma_{10}\rangle$, where $e^{iA}= \langle\Gamma_{00}|\Gamma_{11}\rangle$.

To summarize, MDI QKD with uncharacterized qubit has an advantage that the condition for security is weaker than that of normal MDI QKD's. However, disadvantage is low tolerable bit error rate due to the overestimation for the phase error rate. Here we proposed an improved bound for phase error rate. By simulations we showed the bound considerably improved final key rates.

\section*{Acknowledgement}
This study was supported by Institute for Information and Communications Technology Promotion (IITP) grant funded by the Korea Government (MSIP) (No. R0190-16-2028, Practical and Secure Quantum Key Distribution).


\end{document}